\documentclass[aps,pre,reprint]{revtex4-1}

\usepackage[utf8]{inputenc}
\usepackage{amsmath}
\usepackage{amsfonts}
\usepackage{amssymb}
\usepackage{amsthm}
\usepackage{amsopn}
\usepackage{upgreek}
\usepackage{bm}
\usepackage{epsfig}

\setcitestyle{square,numbers,sort&compress}
\renewcommand{\eqref}[1]{Eq.~(\ref{#1})}

\newcommand{\figref}[1]{Figure~\ref{#1}}
\newcommand{\figsref}[1]{Figures~\ref{#1}}
\newcommand{\refcite}[1]{Ref.~\onlinecite{#1}}

\newcommand{\bens}{\beta\epsilon}
\newcommand{\bess}{\beta\epsilon_{\text{s-s}}}
\newcommand{\becs}{\beta\epsilon_{\text{c-s}}}
\newcommand{\becc}{\beta\epsilon_{\text{c-c}}}

\newcommand{\pM}{p_{\mathcal{M}}}
\newcommand{\M}{\mathcal{M}}
\newcommand{\dM}{\delta\mathcal{M}}
\newcommand{\nsamples}{{n_{\text{samples}}}}
\newcommand{\kB}{k_{\text{B}}}
\newcommand{\sumi}{\!\sum_{i \in \{\text{s},\text{c}\}}\!}
\newcommand{\sumj}{\!\sum_{j \in \{\text{s},\text{c}\}}\!}

\newcommand{\pcc}{p_{\text{c-c}}}
\newcommand{\rhoc}{\rho_{\text{c}}}
\newcommand{\rhos}{\rho_{\text{s}}}
\newcommand{\xc}{x_{\text{c}}}
\newcommand{\xs}{x_{\text{s}}}
\newcommand{\bmuc}{\beta\mu_{\text{c}}}
\newcommand{\bmus}{\beta\mu_{\text{s}}}

\begin{document}

\newcommand{\addresscambridge}{Department of Chemistry, University of Cambridge, Lensfield Road, Cambridge CB2 1EW, United Kingdom}

\title{Oligomers of heat-shock proteins: Structures that don't imply function}
\author{William M.~Jacobs}
\altaffiliation[Present address:~]{Department of Chemistry and Chemical Biology, Harvard University, 12 Oxford Street, Cambridge, MA, 02138, USA}
\email{wjacobs@fas.harvard.edu}
\affiliation{\addresscambridge}
\author{Tuomas~P.J.~Knowles}
\affiliation{\addresscambridge}
\author{Daan~Frenkel}
\email{df246@cam.ac.uk}
\affiliation{\addresscambridge}

\begin{abstract}
Most proteins must remain soluble in the cytosol in order to perform their biological functions.
To protect against undesired protein aggregation, living cells maintain a population of molecular chaperones that ensure the solubility of the proteome.
Here we report simulations of a lattice model of interacting proteins to understand how low concentrations of passive molecular chaperones, such as small heat-shock proteins, suppress thermodynamic instabilities in protein solutions.
Given fixed concentrations of chaperones and client proteins, the solubility of the proteome can be increased by tuning the chaperone--client binding strength.
Surprisingly, we find that the binding strength that optimizes solubility while preventing irreversible chaperone binding also promotes the formation of weakly bound chaperone oligomers, although the presence of these oligomers does not significantly affect the thermodynamic stability of the solution.
Such oligomers are commonly observed in experiments on small heat-shock proteins, but their connection to the biological function of these chaperones has remained unclear.
Our simulations suggest that this clustering may not have any essential biological function, but rather emerges as a natural side-effect of optimizing the thermodynamic stability of the proteome.
\end{abstract}

\maketitle

\section*{Introduction}

Passive molecular chaperones inhibit the aggregation of cytosolic proteins and are thus a nearly ubiquitous component of living cells~\cite{ellis2005chaperone,tyedmers2010cellular,drummond2008mistranslation}.
This class of chaperones comprises clusterin, $\alpha$-crystallins and many other small heat-shock proteins (sHSPs), which promote tolerance to a wide range of cellular stressors such as elevated temperatures and hazardous nonspecific interactions~\cite{humphreys1999clusterin,bakthisaran2015small}.
These chaperones cannot by themselves fold or refold misassembled proteins and do not require ATP to function.
Instead, passive chaperones associate reversibly with aggregation-prone proteins.
Even when present in sub-stoichiometric ratios with their client proteins, sHSPs and similar chaperones are effective at suppressing aggregation and coping with environmental stress~\cite{treweek2003intracellular,carver2003small,hochberg2014structured}.
Yet the mechanism by which this class of chaperones stabilizes the cytosol is not well understood despite significant efforts at determining the structural properties of these molecules.

Here we propose that passive chaperones function by increasing the overall solubility of the proteome.
Through this mechanism, passive chaperones reduce the fraction of toxic oligomers in solution and suppress the nucleation of protein aggregates.
It has recently become apparent that some sHSPs can also interact with protein aggregates in order to curtail further protein deposition~\cite{shammas2011binding,waudby2010interaction,knowles2007kinetics}.
These aggregates are often detrimental to cellular survival, in part because they can sequester other crucial proteins~\cite{bence2001impairment}.
We provide simulation evidence that this effect on the proteome solubility is a generic feature of passive chaperones that associate promiscuously and reversibly with their clients.

There is substantial experimental evidence that passive chaperones interact promiscuously with client proteins in chemical equilibrium.
Both the rate of client aggregation and the fraction of chaperones associated with insoluble proteins are concentration-dependent~\cite{ellis2005chaperone,drummond2008mistranslation}.
Furthermore, chaperone binding responds directly to increases in the available client binding surfaces, including hydrophobic regions of destabilized clients that are only transiently exposed~\cite{mchaourab2009structure}.
The binding of passive chaperones often modifies the size and structure of amorphous aggregates, leading to smaller soluble clusters in which the putative chaperone binding sites are associated with the hydrophobic interfaces of the client proteins~\cite{haslbeck1999hsp26,basha2012small,haslbeck2005some}.
On the basis of these dynamic chaperone--client aggregates, previous studies have suggested that such aggregates might serve as a relatively inert depot of misfolded proteins during cellular stress~\cite{tyedmers2010cellular,arrasate2004inclusion,tanaka2004aggresomes,hartl2011molecular,park2013polyq}.

However, client proteins are not the only substrates to which passive chaperones bind: these chaperones are commonly found in chaperone-only oligomers both \textit{in vitro} and \textit{in vivo}~\cite{carver2003small,haslbeck1999hsp26,haslbeck2005some,jakob1993small,theriault2004essential,van2002structure,robertson2010small,basha2012small}.
Recent experiments indicate that these dynamic oligomers are also under thermodynamic control~\cite{hochberg2014dynamical,basha2012small,haslbeck2005some,baldwin2011quaternary} and vary with the experimental conditions, such as the temperature and the ionic strength of the solution~\cite{hochberg2014dynamical,sun2005small,fu2003small}.
Because this tendency to form oligomers is highly conserved across the family of sHSPs and similar molecular chaperones, it has long been recognized that dynamic fluctuations in the oligomeric state play an important role in the organization of many passive chaperones~\cite{carver2003small,benesch2008small,stengel2010quaternary,hochberg2014dynamical}.
At present, however, it is unclear whether the formation of chaperone oligomers is a key \textit{functional} event.
In fact, there is considerable evidence to the contrary: experiments have shown that mutations and post-translational modifications that alter the tendency of chaperones to form oligomers do not necessarily affect their function~\cite{sun2005small,hayes2009phosphorylation,feil2001novel,haslbeck2004domain,aquilina2005subunit}.
These observations raise the question of how, if at all, the presence of chaperone oligomers contributes to their ability to solubilize aggregation-prone proteins \textit{in vivo}.
Here we show that both the function and oligomerization of passive molecular chaperones can be explained by identifying the optimal conditions for a thermodynamically stable solution of chaperones and aggregation-prone proteins.
Our results suggest that low concentrations of promiscuous chaperones are a generic means of stabilizing a biological mixture with respect to a variety of nonfunctional interactions.

\section*{Results}

To understand how passive molecular chaperones affect the thermodynamic stability of a protein solution, we consider a minimal model of two species in solution: an aggregation-prone protein and a simple molecular chaperone.
Aggregation of the client proteins is primarily driven by highly directional interactions.
These interactions are mediated by `patches,' which represent primarily hydrophobic regions that are commonly involved in both functional and aberrant protein--protein interactions.
Chaperone--client recognition is also driven by these directional associations between chaperone monomers and the exposed patches of client monomers.
Both the chaperone and client proteins may also associate via weak nonspecific interactions, which we assume to be averaged over the relative orientations of the monomers.
These pairwise isotropic interactions account for transient associations between proteins in a crowded environment~\cite{jacobs2014phase,jacobs2013predicting}.
We do not explicitly model the overwhelming majority of proteins that may also experience this weak nonspecific interaction but are not prone to aggregation via directional interactions, as this simplification does not qualitatively affect our analysis.

\subsection*{Lattice model of a chaperone--client mixture}

\begin{figure}[t]
  \centering
  \includegraphics{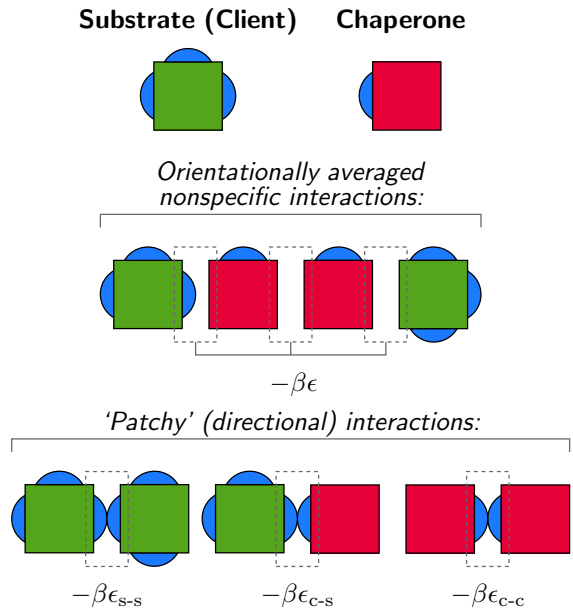}
  \caption{A minimal model of an associating fluid of passive chaperones and aggregation-prone client proteins.  Chaperone and client monomers interact via nearest-neighbor interactions on a three-dimensional cubic lattice.  Orientationally averaged nonspecific interactions may be either attractive or repulsive.  Directional interactions between specific binding sites, indicated by blue patches, depend on the relative orientations of the monomers and are always attractive.}
  \label{fig:model}
\end{figure}

In protein solutions under physiological conditions, the interactions between proteins are short-ranged in comparison to the size of the monomers, since the high ionic strength characteristic of physiological media leads to an effective screening of electrostatic interactions~\cite{gunton2007protein}.
We therefore choose to model protein interactions through nearest-neighbor contacts on a three-dimensional lattice, where unoccupied lattice sites represent an implicit solvent.
Monomers interact if they reside on adjacent lattice sites, and they are free to rotate and to move among lattice sites in accordance with the equilibrium Boltzmann distribution.
We assume that each protein exists in a single coarse-grained conformation and that the interactions between proteins are determined by effective binding free energies (\figref{fig:model}).
This coarse-graining of the internal degrees of freedom allows us to capture the effects of the intermolecular forces in a reduced set of parameters and is particularly suitable for both globular proteins in near-native states and misassembled proteins with exposed hydrophobic regions.
All monomers on adjacent lattice sites experience an orientationally averaged nonspecific interaction, which is assigned a dimensionless free energy of $-\bens$.
(Interaction energies are expressed in thermal units: ${\beta^{-1} \equiv \kB T}$, where $\kB$ is the Boltzmann constant and $T$ is the absolute temperature.)

Because aggregation-prone proteins are likely to participate in directional protein--protein interactions via multiple binding sites~\cite{ellis2005chaperone,pastore2012two,cumberworth2013promiscuity,de2012evolutionary}, which also promote interactions with sHSPs~\cite{carver2002interaction,fu2014multilevel}, we choose a client model with three patches that is susceptible to aggregation by means of directional interactions alone (\figref{fig:model}).
The directional interactions between client monomers are assigned an attractive free energy of $-\bess$.
These interactions are chosen to be strong enough to form insoluble client aggregates in the absence of both chaperones and additional nonspecific interactions~\cite{jacobs2014phase}.

A minimal model of a passive chaperone must be capable of binding exposed patches on the client monomers.
Here we assume that the chaperone monomers have a single binding site and that the interaction free energy between chaperone and client patches is $-\becs$ (\figref{fig:model}).
While this assumption is clearly a simplification of the structure of passive chaperones, which may interact with diverse clients via different binding sites, this representation captures the passivation of interactive client binding sites through the burial of hydrophobic surfaces.
Most importantly, this representation has the physical features that are necessary to capture the qualitative effects of passive chaperones on the thermodynamics of a complex fluid.
Because passive chaperones are known to function at low concentrations, we assume that there are always fewer chaperones than client binding sites.
In what follows, the relative amounts of the chaperone and client monomers in solution are indicated by $\xc$ and $\xs$, respectively, such that ${\xc + \xs = 1}$.

We have used Monte Carlo simulations and finite-size-scaling techniques to calculate the miscibility limit of this model, i.e., the point at which the chaperone--client mixture becomes unstable with respect to aggregation and/or demixing (see Methods).
This miscibility limit coincides with a thermodynamic instability, where small, spontaneous fluctuations are sufficient to establish long-ranged spatial heterogeneity in an initially well-mixed solution.
In a protein solution, a thermodynamic instability may have contributions from directional interactions, which cause the polymerization and demixing of the strongly interacting species, as well as orientationally averaged interactions, which drive the formation of thermodynamic phases with differing protein densities~\cite{jacobs2014phase}.
Strong directional interactions between the client proteins can thus lead to the formation of disordered aggregates and an accompanying loss of solubility.

\subsection*{Passive chaperones enhance the thermodynamic stability of a protein solution}

\begin{figure}[t]
  \centering
  \includegraphics{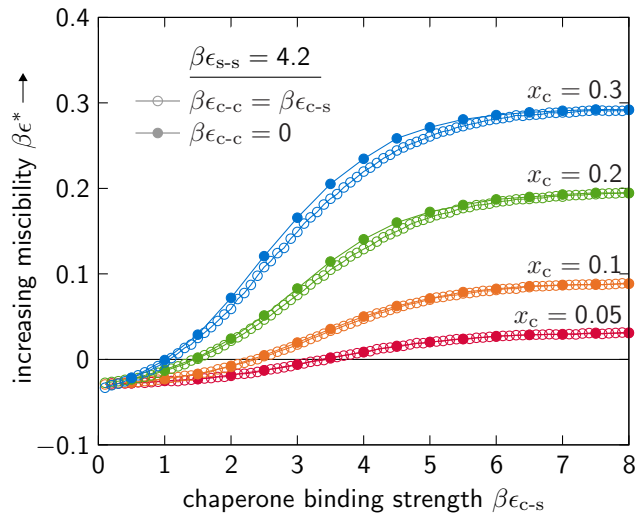}
  \caption{The miscibility limit, $\bens^*\!$, of a chaperone--client mixture depends on the chaperone--client binding strength, $\becs$, and the chaperone stoichiometric fraction, $\xc$.  The chaperone--chaperone interactions have only a minor effect on $\bens^*\!$: the miscibility limits of solutions with promiscuous chaperone interfaces, for which ${\becc = \becs}$, are indicated by open circles, while the miscibility limits of solutions in which chaperone--chaperone interactions are prevented are indicated by closed circles.  The client--client interaction strength, $\bess$, is sufficient to drive the aggregation of clients in the absence of chaperones.}
  \label{fig:miscibility}
\end{figure}

As expected, the presence of chaperones inhibits the formation of client oligomers by competing for binding to patches on the client monomers.
However, this passivation of directional interactions is not the only effect of chaperone binding: the interactions between chaperones and client proteins simultaneously increase the strength of the orientationally averaged nonspecific interactions that the solution can tolerate while remaining thermodynamically stable.
This effect can be seen in \figref{fig:miscibility}, which shows the miscibility limit, $\bens^*\!$, at which insoluble aggregates first appear in the solution.
When the strength of the orientationally averaged nonspecific interactions increases beyond this limit, i.e., ${\bens > \bens^*\!}$, the solution becomes unstable with respect to small fluctuations in the protein concentrations.
Increasing the strength of these nonspecific interactions can thus cause the solution to become unstable without altering the strength of the directional interactions that drive the polymerization of the client monomers.
Our calculations show that passive chaperones dramatically affect the miscibility limit by inhibiting polymerization and solubilizing transient clusters of client proteins, despite the fact that there are far fewer chaperones than there are client binding sites.

In the absence of chaperones, i.e., $\xc \rightarrow 0$, the solution is unstable due to the strong directional interactions between the aggregating client monomers.
In this case, $\bens^*$ is negative, indicating that a solution of sufficiently concentrated client proteins in a well-screened solvent will form insoluble aggregates.
It is important to note that even when ${\becs = 0}$, the chaperones still interact nonspecifically with the client monomers through the orientationally averaged interaction $\bens$.
Here we find that the addition of such `inert' chaperones has a negligible effect on the miscibility limit relative to a client-only solution.
This observation also implies that the majority of cytosolic proteins that are not aggregation-prone do not significantly affect the miscibility limit when the dominant instability is driven by strong directional interactions.

Our calculations further indicate that the thermodynamic forces driving these instabilities are qualitatively different in solutions with weakly and strongly binding chaperones.
In the case of weakly binding chaperones (${\becs \ll \bess}$), the solution demixes into client-enriched and client-depleted phases primarily as a result of directional interactions.
Insoluble client aggregates recruit monomers via the formation of directional contacts and exchange small oligomers with the coexisting solution.
With strongly binding chaperones (${\becs \gg \bess}$), the solution forms a high-density condensate consisting of both chaperones and client proteins bound by nonspecific interactions.
Under these conditions, the proteins in both the soluble and insoluble phases exist as amorphous clusters that decrease in size as the stoichiometric ratio ${\xc / \xs}$ is increased.
The introduction of strongly binding chaperones, even in low concentrations, significantly increases the solution miscibility limit towards the theoretical maximum for this model, ${\bens^*_{\text{max}} \simeq 0.87}$.

\subsection*{Chaperone--chaperone interactions do not significantly affect the miscibility limit}

Because a chaperone that interacts with a variety of clients is very likely to engage in promiscuous interactions, it is reasonable to assume that chaperones do not distinguish among the various hydrophobic surfaces in solution.
The strength of interactions between chaperone binding sites is likely to be similar to the strength of interactions between chaperones and clients, and thus it is natural to assume that ${\becc = \becs}$.
However, if we instead prevent chaperone--chaperone binding by setting ${\becc = 0}$, we find that the effect on the solution miscibility limit is negligible (\figref{fig:miscibility}).

Since the parameter $\becc$ directly controls the probability of chaperone dimerization, our calculations suggest that the formation of chaperone oligomers has a very minor effect on chaperone function.
Experimentally, the relationship between oligomerization and chaperone function has been probed by modifying or truncating sHSPs~\cite{sun2005small,feil2001novel,haslbeck2004domain,aquilina2005subunit}.
The available experimental evidence indicates that alterations to the putative client binding sites on sHSPs affect the oligomer equilibria and the functionality of the chaperones independently, in qualitative agreement with the predictions of this model.

\subsection*{A biological fitness function suggests conditions for optimized chaperone operation}

Putting these results into context, we now ask, ``Is there an optimal chaperone--client binding strength for a biological mixture?''
\figref{fig:miscibility} shows that strongly binding chaperones are best suited for increasing the miscibility limit.
In this case, producing more chaperones (or reducing the total concentration of aggregation-prone clients) increases $\bens^*$ in an approximately linear relationship, allowing an organism to respond effectively to an increase in nonspecific interactions.
Nevertheless, strong promiscuous interactions come at a cost: nearly irreversible binding between a chaperone and any available association site, including other proteins that are not explicitly modeled in our simulations, sequesters both interfaces, thereby preventing their participation in further functional interactions.
The optimal chaperone binding strength must balance these competing requirements for solution stability and reversible binding.

Despite the complexity of naturally occurring protein solutions, we can predict the optimal chaperone binding strength by considering a generic fitness function, which quantifies the trade-offs in biological costs and benefits.
This fitness should be maximized for optimal biological function.
For the present model, the fitness $\mathcal{F}$ is a function of the miscibility limit, $\bens^*\!$, as well as two biological costs that depend on the variables $\becs$ and $\xc$.
In the absence of any deleterious effects of chaperone action, increasing the solubility of the proteome must be beneficial, and thus $\mathcal{F}$ should be an increasing function of ${\bens^*\!}$.
However, one potential cost of chaperone action arises from the sequestration of functional proteins (which are not explicitly modeled in our simulations but must be present in a naturally occurring protein solution) due to the promiscuous binding of chaperones.
Another potential cost is associated with the production of chaperone molecules.
These costs imply that the fitness function ${\mathcal{F}[\bens^*(\becs,\xc), \becs, \xc]}$ should satisfy both ${\left.\partial \mathcal{F} / \partial \becs \right|_{\bens^*} < 0}$ and ${\left.\partial \mathcal{F} / \partial \xc \right|_{\bens^*} < 0}$, respectively, when the miscibility limit $\bens^*$ is held constant.
Taking the total derivative of $\mathcal{F}$ with respect to both $\becs$ and $\xc$, we find that this fitness function is maximized where
\begin{equation}
  \nonumber
  \frac{\partial \mathcal{\bens^*}}{\partial \becs} = \frac{-\left.\displaystyle\frac{\partial \mathcal{F}}{\partial \becs} \right|_{\bens^*}}{\displaystyle\frac{\partial \mathcal{F}}{\partial \bens^*}}
  \quad
  \text{and}
  \quad
  \frac{\partial \mathcal{\bens^*}}{\partial \xc} = \frac{-\displaystyle\left.\frac{\partial \mathcal{F}}{\partial \xc} \right|_{\bens^*}}{\displaystyle\frac{\partial \mathcal{F}}{\partial \bens^*}}.
\end{equation}
All partial derivatives of $\mathcal{F}$ depend on the precise nature of the biological system and thus cannot be determined precisely.
Nevertheless, it is reasonable to assume that the cost derivatives of $\mathcal{F}$ are approximately constant: at low chaperone concentrations, the law of mass action implies that the cost due to reversible, promiscuous binding is approximately linear in both $\xc$ and $\becs$, while the total cost associated with the production of chaperone molecules is also proportional to their concentration.
We can therefore interpret the ratios of the derivatives in each of the above equations as the importance of each cost relative to the benefit of stabilizing the protein solution.
Assuming that promiscuous chaperone binding and chaperone production are indeed significant biological costs, then these equations imply that we should seek to optimize the fitness by maximizing the response functions ${\partial \bens^* / \partial \becs}$ and ${\partial \bens^* / \partial \xc}$.

More intuitively, maximizing these response functions directs the optimal chaperone design towards the region of parameter space in which the solution miscibility limit is most sensitive to small increases in either the chaperone--client binding strength or the number of chaperone molecules in solution.
The first condition, ${\partial \bens^* / \partial \becs}$, biases the optimal chaperone design away from values of $\becs$ for which the miscibility limit increases asymptotically, thus discriminating against excessively strong binding between chaperones and clients.
The second condition, ${\partial \bens^* / \partial \xc}$, requires that the miscibility limit be sensitive to changes in the chaperone stoichiometric fraction.

\begin{figure*}[t]
  \centering
  \includegraphics{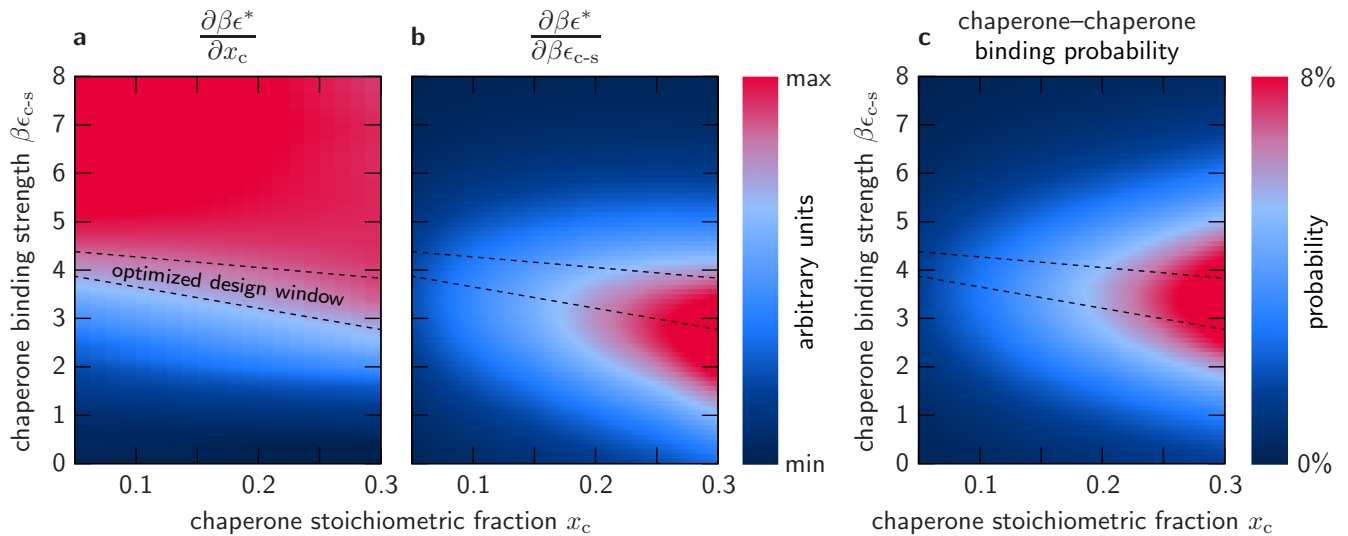}
  \caption{The optimized design window for a passive molecular chaperone coincides with the conditions under which chaperone oligomerization is most probable.  (a) The response in the solution miscibility limit to an increase in the chaperone stoichiometric fraction.  The approximately linear response regime for strong binding chaperones is indicated in red.  (b) The response in the solution miscibility limit to an increase in the chaperone--client binding strength.  (c) The probability that a chaperone binding interface is bound to another chaperone monomer.  For a given stoichiometric ratio of chaperones and clients, this probability is greatest in approximately the same design window in which both $\partial \bens^* / \partial \xc$ and $\partial \bens^* / \partial \becs$ are simultaneously maximized.}
  \label{fig:response}
\end{figure*}

Our calculations show that it is indeed possible to satisfy both conditions simultaneously.
In \figsref{fig:response}a and \ref{fig:response}b, we plot the calculated response functions ${\partial \bens^* / \partial \becs}$ and ${\partial \bens^* / \partial \xc}$, in dimensionless units, as functions of the chaperone stoichiometric fraction and the chaperone binding strength.
We identify a `design window' for optimal chaperone operation by finding the approximate range of chaperone binding strengths over which both response functions are maximized given a fixed chaperone stoichiometric fraction.
The region of parameter space in which both response functions can be maximized is relatively narrow, suggesting that optimized passive chaperones should have tightly constrained binding strengths.
We further find that the optimal range of chaperone binding strengths is only weakly dependent on the chaperone stoichiometric fraction.
This observation implies that chaperones with fixed binding strengths can operate close to optimality over a wide range of sub-stoichiometric concentrations.
We also note that these protein--protein interaction free energies are in the physical range of a few $\kB T$.
The optimal chaperone binding strength is generally weaker than the client--client interactions, indicating that the chaperones need not out-compete the aggregation-prone clients for association with exposed binding interfaces.

\subsection*{The optimal chaperone--client binding strength promotes chaperone oligomerization}

Remarkably, our simulations reveal that the probability of finding chaperone oligomers is also highest in the region of parameter space where the optimal design conditions for chaperone activity are satisfied.
In \figref{fig:response}c, we plot the probability of chaperone--chaperone binding at the miscibility limit, assuming that ${\becc = \becs}$.
We find that this probability is maximal in the window of optimal chaperone binding strength over the complete range of simulated chaperone stoichiometric fractions.
Under these conditions, a significant fraction of the chaperone binding sites are not associated with the aggregation-prone interfaces on the client proteins, but are rather buried in chaperone-only oligomers.
This fraction may be even higher in the miscible fluid or in the presence of client proteins that are less prone to aggregation due to weaker directional interactions.

These calculations provide further evidence that the assembly of chaperone oligomers does not play a functional role.
Although the choice of $\becc$ affects the magnitude of the effect shown in \figref{fig:response}c, we emphasize that simply allowing chaperone--chaperone binding does not imply that chaperone-only oligomers will be observed at the miscibility limit: there is a large region of parameter space over which this probability is very small.
Furthermore, the results presented in \figref{fig:response} are qualitatively unchanged for all reasonable choices of $\becc$, i.e., ${0 < \becc \lesssim \becs}$.
Our simulations thus indicate that the ability to assemble chaperone oligomers affects neither the anti-aggregation function of the chaperones nor their adherence to the proposed design constraints.

\section*{Discussion}

We have shown that a simple model of chaperone--client mixtures reveals two generic and unexpected features of passive molecular chaperones.
First, chaperone--chaperone interactions only marginally affect the stability of a protein solution in which strong directional interactions drive the aggregation of client proteins.
Second, promiscuous passive chaperones tend to assemble chaperone oligomers under conditions where the chaperone--client binding strength balances the requirement for proteome stability with the need to avoid irreversible binding.
Taken together, these results suggest that the assembly of oligomers of passive molecular chaperones is not  an essential functional event for stabilizing a protein solution.
Instead, this behavior emerges as a side-effect of operating under thermodynamically optimal conditions.

To arrive at this conclusion, we have proposed that passive chaperones perform their anti-aggregation function by increasing the miscibility limit of a protein solution.
Through this mechanism, passive chaperones inhibit the sequestration of functional proteins and increase the thermodynamic stability of a biological mixture with respect to random nonspecific interactions.
Our simulations demonstrate that this mechanism is physically plausible even when the aggregation-prone client proteins greatly outnumber the chaperones.
We emphasize that only the ratio of chaperone molecules to client binding interfaces, not the total concentration of chaperones in solution, is relevant for chaperone function.
In all cases considered here, the stoichiometric fraction of chaperones is much lower than ${\xc = 0.75}$, the fraction that would be required to passivate all binding sites on the three-patch client monomers in the solution.
The fact that the chaperones that fulfill this anti-aggregation function are highly conserved in both lower and higher organisms suggests that there is a strong evolutionary pressure to perform this role in an optimized fashion.
Our calculations indicate that the range of suitable chaperone binding strengths is indeed narrow and that the principles for an optimal design emerge from thermodynamic arguments.

The generality of the present model suggests that the assembly of chaperone-only oligomers would not be affected by introducing additional detail in an off-lattice model.
Such an extension, however, would allow a much wider variety of chaperone oligomers to be observed.
The significant coarse-graining involved in the development of the present model and the high symmetry imposed by the lattice do not permit the reproduction of many structural features or the precise oligomeric distributions of specific passive chaperones.
For instance, all three domains of $\alpha${B}-crystallin are believed to be involved in the assembly of higher-order oligomers~\cite{baldwin2011quaternary}, while the chaperones in the present model may only form dimers through client-binding interfaces.
Nevertheless, such detailed molecular interactions are unlikely to affect the physical mechanism by which passive chaperones suppress aggregation.

Most importantly, the simplicity of this model allows us to make generic predictions about the thermodynamics of passive molecular chaperones.
Regardless of the molecular-level details, the critical behavior of a fluid with short-ranged interactions falls within the Ising universality class, which is also known to describe phase separation in globular protein solutions~\cite{thomson1987binary,schurtenberger1989observation,taratuta1990liquid,broide1991binary,wang2013phase}.
In the vicinity of the miscibility limit, fluctuations in both the protein density and the intermolecular contacts within aggregates are significant, and a broad distribution of cluster sizes is observed at equilibrium.
Our proposed mechanism therefore supports the assertion that subunit exchange is essential for the function of sHSPs and related chaperones~\cite{carver2003small,stengel2010quaternary,baldwin2012probing,bova1997subunit,humphreys1999clusterin}.
Even if the aggregates are not fully equilibrated due to slow kinetics, the large concentration fluctuations in the vicinity of a metastable critical point are likely to enhance the formation of gel-like aggregates~\cite{li2012phase} or the nucleation of aggregated phases~\cite{ten1997enhancement,vekilov2012phase}.
For example, recent simulations have shown that clustering through nonspecific interactions plays an important role in the kinetics of amyloid fibril nucleation~\cite{saric2014crucial}.

\section*{Conclusions}

We have presented a minimal model of a mixture of passive molecular chaperones and aggregation-prone proteins.
By calculating the limit of thermodynamic stability in this model protein solution, we have shown how passive chaperones that are expressed in sub-stoichiometric ratios with their clients can substantially suppress aggregation.
We have further argued that the biological costs associated with chaperone production and promiscuous, irreversible binding significantly constrain the optimal design of an effective passive chaperone.
We find that if passive chaperones interact promiscuously with exposed hydrophobic surfaces, then the assembly of chaperone oligomers emerges as a nonfunctional side-effect of this thermodynamically optimal design.
Because of the generality of the model, these conclusions are relevant to a broad class of molecular chaperones.
Fully atomistic simulations could provide further information on the parameters governing the interaction strengths between chaperones and their aggregation-prone targets as well as between the passive chaperones themselves.
Such simulations could therefore provide a means of transferring the general thermodynamic principles uncovered by the coarse-grained simulations presented here to detailed models of specific chaperone--client mixtures.

\section*{Methods}

In the lattice model considered here, the limit of thermodynamic stability of a well-mixed solution is encountered at the critical surface for phase separation.
In what follows, we describe the Monte Carlo simulations and finite-size-scaling theory used to calculate points on this critical surface.
Our approach is a generalization of the computational strategy described in detail in \refcite{jacobs2014phase}.

In general, the critical surface of a multicomponent mixture has dimension ${d - 2}$, where $d$ is the total number of independent thermodynamic fields~\cite{griffiths1970critical}.
The independent thermodynamic fields in the present model are the dimensionless chemical potentials of both the chaperones and the clients, $\bmuc$ and $\bmus$, respectively, as well as the dimensionless interaction energies: $\bens$, $\bess$, $\becs$ and $\becc$.
The relevant critical surface in this model is thus a 4-dimensional surface.

We perform biased grand-canonical Monte Carlo simulations, as described in \refcite{jacobs2014phase}, to collect statistically independent lattice configurations near the critical surface.
We use a ${L \times L \times L}$ cubic lattice with periodic boundary conditions and set ${L = 12}$ so that all simulations are carried out in the scaling regime.
We then apply the finite-size-scaling theory of Wilding and Bruce~\cite{bruce1992scaling,wilding1997simulation} to solve self-consistently for the critical order parameter, $\hat\M$, and the critical orientationally averaged nonspecific energy, $\bens^*\!$, at fixed values of $\bess$, $\becs$, $\becc$ and $\xc$.
In order to determine each critical point plotted in \figref{fig:miscibility}, we approximate the marginal probability distribution $p(\M)$ from the grand-canonical samples and then tune this distribution in order to match the known distribution of the critical ordering operator in the three-dimensional Ising universality class, $\pM$.
This computational procedure is described below.

In a two-solute solution, with two independent dimensionless chemical potentials $\bmuc$ and $\bmus$, the critical order parameter must account for fluctuations in the number densities of both the client and chaperone monomers, $\rhos$ and $\rhoc$, respectively, as well as fluctuations in the internal energy density, $u$.
The critical fluctuations in the number densities can be described by the vector $\hat\nu$, which indicates the difference in compositions of the two incipient phases~\cite{jacobs2013predicting}.
We therefore define $\hat\M$ to be the linear combination
\begin{equation}
  \label{eq:M}
  \hat\M \equiv \nu_{\text{s}} \hat\rhos + \nu_{\text{c}} \hat\rhoc - s \hat{u},
\end{equation}
where both $\hat\nu$ and the field-mixing parameter $s$ must be determined self-consistently.
The grand-canonical distribution of $\M$ is constructed from the simulation data according to
\begin{equation}
  \label{eq:pgc_bin}
  p_{\text{gc},k}^{(\M)} \equiv \Lambda\! \sum_v w_v \bm{1}\! \left\{ \dM_k \le \left[ (\rhos,\rhoc,u)_v \!\cdot\! \hat{\M} \right] \! < \dM_{k+1} \right\}\!,
\end{equation}
where the index $v$ runs over all independent samples and ${\bm{1}\{ \cdot \}}$ is the indicator function.
Each sample has a statistical weight ${w_v}$ in the grand-canonical ensemble that depends on the values of the thermodynamic fields \cite{jacobs2014phase}.
The system-dependent scaling constant $\Lambda$ must be determined self-consistently.
The bin size is chosen such that ${(\dM_{k+1} - \dM_k) = L^{-3}}$, where ${\dM \equiv \Lambda (\M - \M^*)}$ and $\M^*$ is the ensemble-averaged mean value of $\M$.

We then construct a $\chi^2$-function that seeks to minimize the difference between the observed distribution of $\M$ and the universal distribution, $\pM$, while obeying the imposed composition constraint:
\begin{eqnarray}
  \label{eq:chi2}
  \chi^2 &\equiv& \sum_k \frac{\left[ p_{\text{gc},k}^{(\M)}(\beta\vec{f}\,) - \pM(\dM_k/\Lambda) \right]^2}{\sigma_k^2} + \\
  &\quad& \sumi\!\! \frac{\left( \left\langle \rho_i(\beta\vec{f}\,) \right\rangle / \,\sumj \left\langle \rho_j(\beta\vec{f}\,) \right\rangle - x_i \right)^2}{\sigma_i^2}, \nonumber
\end{eqnarray}
where ${\beta\vec{f} \equiv (\bens, \bess, \becs, \becc, \bmus, \bmuc)}$ and the index $k$ runs over all bins.
In the second term, ${\langle \rho_i \rangle}$ indicates the ensemble-averaged number density of component $i$.
We estimate the error in the sampled distribution of $\M$ to be
\begin{equation}
  \label{eq:pgc_bin_error}
  \sigma_k^2 = \frac{\left( \sum_v w_v^2 \bm{1}_{k,v} \right) - \left( \sum_v w_v \bm{1}_{k,v} \right)^2 / \nsamples} {\sum_v w_v},
\end{equation}
where $\bm{1}_{k,v}$ is the indicator function written out explicitly in \eqref{eq:pgc_bin}, and we estimate the error in the observed composition at the critical point to be
\begin{equation}
  \label{eq:pgc_composition_error}
  \sigma_i^2 = \frac{1}{\phi^2} \!\left[ \sum_{j,k \in \{\text{s},\text{c}\}} \!\!\! \left( \delta_{ij} - \frac{\rho_i}{\phi} \right) \langle \delta\rho_j\delta\rho_k \rangle \left( \delta_{ik} - \frac{\rho_i}{\phi} \right) \right]\!,
\end{equation}
where ${\langle \delta\rho_j\delta\rho_k \rangle \equiv \langle \rho_j \rho_k \rangle - \langle \rho_j \rangle \langle \rho_k \rangle}$, ${\phi \equiv \sumj \rho_j}$ and $\delta_{ij}$ is the Kronecker delta.

Finally, we calculate the probability of chaperone dimerization, $\langle \pcc \rangle$, directly from the simulation data according to the definition
\begin{equation}
  \langle \pcc \rangle^* \equiv \left\langle \frac{2 n_{\text{cc}}}{N_{\text{c}}} \right\rangle^*,
\end{equation}
where $n_{\text{cc}}$ is the number of chaperone--chaperone patch contacts and $N_{\text{c}}$ is the total number of chaperone monomers on the lattice.
In this definition, $\langle \cdot \rangle^*$ indicates a grand-canonical average obtained at the critical point with the specified chemical potentials and directional interaction energies.

\end{document}